# Mechanisms of Zener diode and ohmic contact resulting from a heavily doped p-n junction: A novel explanation in terms of impurity-photovoltaic-effect resulting from infrared self-emission at room-temperature

Jianming Li

Institute of Semiconductors, Chinese Academy of Sciences, A35 Qinghua East Road, Haidian District, Beijing 100083, P. R. China

**ABSTRACT**

In textbooks, a quantum-mechanical tunneling mechanism based on electron-wavefunction is used to explain Zener diode as well as ohmic contact resulting from a heavily doped p-n junction. In this study, an attempt in particle description is made to explain these devices. Any object at room-temperature emits infrared (IR) photons. The process of heavy doping can cause many defects. The self-absorption of the IR emission could be achieved through sub-band-gap excitations due to the defects, creating carriers. In a heavily doped p-n junction device, some of the IR-generated carriers diffuse into the junction which has a built-in field. The built-in field sweeps out the carriers, creating IR photocurrent. The reverse leakage current of the p-n junction is regarded as the IR photocurrent according to a precedent where impurity-photovoltaic-effect is used to explain circuit devices (*arXiv*:2510.18226). Heavy doping results in not only a great many IR-generated carriers but also an avalanche effect at low voltage, producing Zener breakdown. Additionally, the impurity-photovoltaic-effect can make a reasonable contribution to explaining the temperature dependence of Zener breakdown. If the doping is very heavy, the heavily doped p-n junction can be altered to create an ohmic contact, which can also be explained in terms of impurity-photovoltaic-effect. Not only quantum mechanics, but also this photovoltaic mechanism in particle description can explain the above devices. Considering the wave-particle duality, the combination of quantum-mechanical and photovoltaic mechanisms is suggested to explain these devices. Each mechanism plays a big or small role in the explanation.

## Ⅰ. INTRODUCTION

Diodes were discovered about a hundred years ago. At a breakdown voltage, the reverse current of a diode suddenly rises from an increasing reverse-bias voltage in which a very low dynamic resistance is present [1]. As shown in Fig. 1, Zener breakdown exists in a diode under reverse bias voltage (Vz), which was postulated by Zener in 1934 [2].

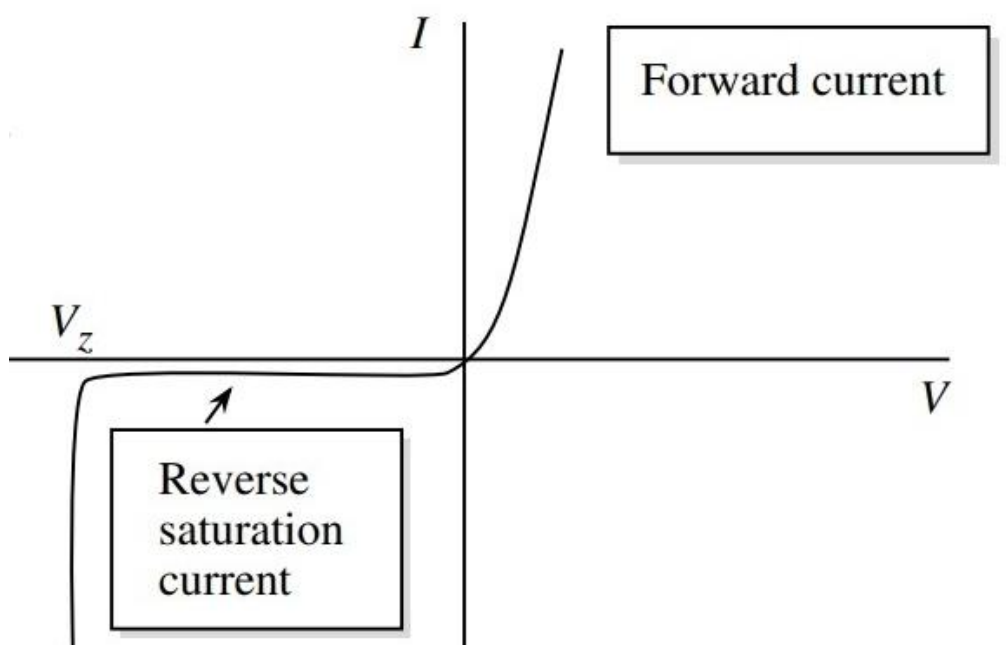

Fig. 1. The current-voltage characteristics of a p-n junction diode. This schematic shows that an abrupt onset of reverse breakdown current occurs as applied voltage approaches breakdown value (Vz).

Zener diodes are one of the semiconductor diodes used in our daily life. A Zener diode has a well-controlled breakdown voltage and is used as a voltage regulator [3]. Zener diodes are a class of semiconductor diodes widely used for protection circuits, constant voltage circuits, level shift circuits, surge clippers etc.

Under conditions of heavy doping, p-n junctions can be altered to Zener diodes or even ohmic contacts. In a tandem multi-junction solar cell, a reversed p–n junction exists between two cell p-n junctions. A challenge for tandem cells has been the development of the effectively electrical connection between the cell p-n junctions. Under conditions of very heavy doping, the reversed p-n junction can be altered to create an ohmic contact.

In textbooks, the theoretical understanding for both Zener diode and ohmic contact is quantum-mechanical tunneling based on electron-wavefunction [4]. Tunneling transmission is an example of wave-particle duality. In quantum mechanics, tunnelling is attributed to an electron wavefunction that extends into a potential barrier. As claimed in a textbook, “tunneling cannot be explained in the particle description, it requires the wave description” [5].

An attempt has been made to explain the ohmic contact of metal-semiconductor by using impurity-photovoltaic-effect in particle description [6]. In this study, the ohmic contact resulting from a heavily doped p-n junction is also explained in terms of impurity-photovoltaic-effect.

## Ⅱ. SUPPOSED MECHANISM

Since 2004, the impurity-photovoltaic-effect resulting from infrared self-emission at room-temperature has been proposed to create photocurrent in a p-n junction diode with unavoidable intrinsic background-defects [7,8].

There is already a precedent where impurity-photovoltaic-effect is used to explain circuit devices [9]. Following the precedent, a mechanism based on impurity-photovoltaic-effect is supposed as follows.

As is well known, any object at room-temperature emits infrared (IR) photons due to blackbody radiation. Some research results [10,11] support that semiconductor devices at room temperature emit IR photons. It is also known that heavy doping process can induce a lot of defects such as vacancies, interstitials, and imperfections associated with impurities, etc.

In a heavily doped p-n junction diode, the defects create a lot of intermediate levels (bandgap states) in forbidden energy band-gap. The self-absorption of the IR emission within the diode could be achieved through the sub-band-gap excitations shown in Fig. 2, which produces a lot of electron–hole pairs.

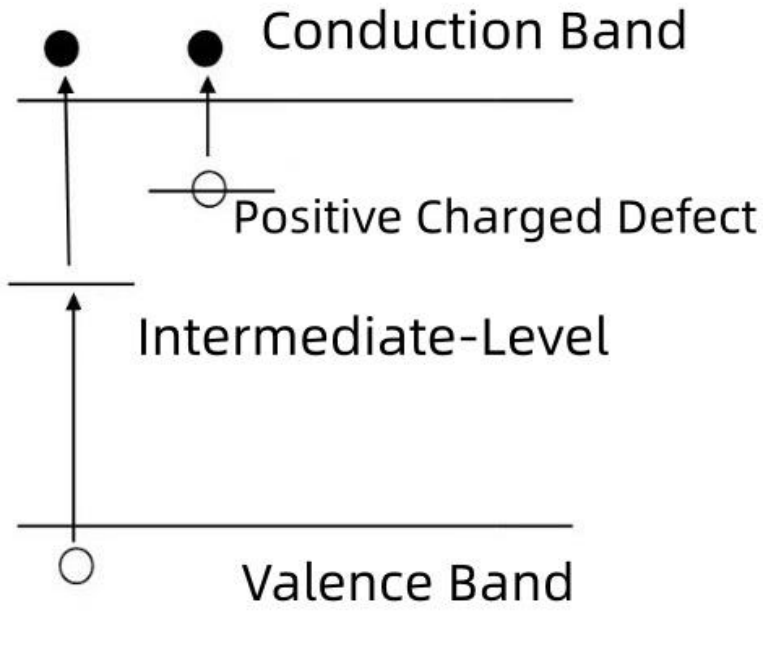

Fig. 2. The sub-band-gap excitation mechanism via the intermediate levels within forbidden gap.

As shown in Fig. 3(a), a depletion layer (a space-charge region with built-in electric field) exists in a p-n junction. Some of the IR-generated electrons and holes diffuse into the depletion layer where the electrons and holes drift in opposite directions due to the built-in field. Therefore, the p-n junction separates the IR-generated electrons and holes, producing IR photocurrent. A series of studies have shown that the reverse leakage current of a p-n junction is regarded as the IR photocurrent due to unavoidable intrinsic background-defects [7-9].

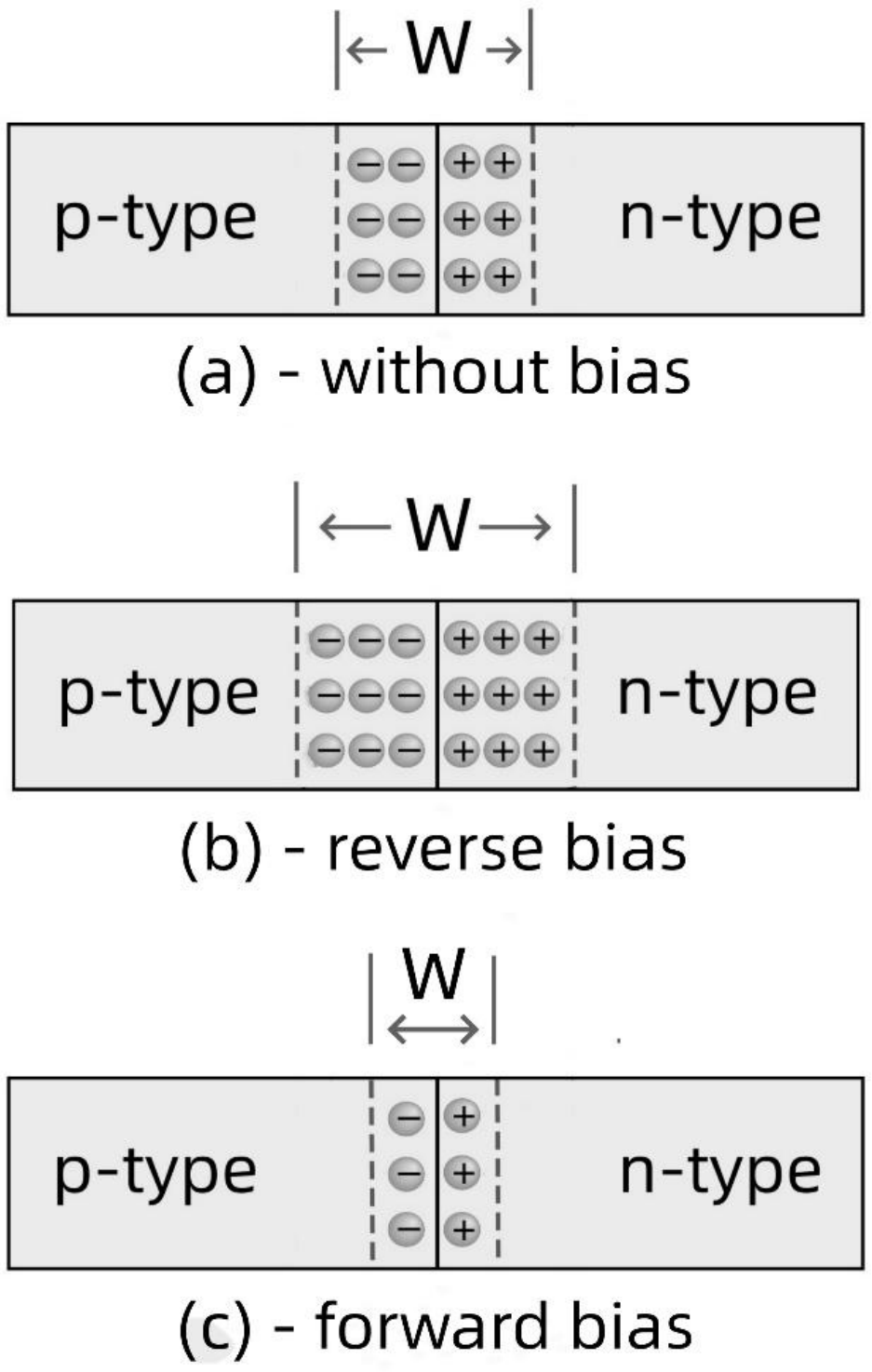

Fig. 3. A schematic of a p-n junction. (a) In thermodynamic equilibrium (without bias), (b) reverse bias, and (c) forward bias. W is depletion layer width.

## Ⅲ. DISCUSSION

Compared to an ordinary diode, a Zener diode has higher doping concentration. Here, let us consider a p-n junction diode in which both p and n sides are heavily doped. Heavy doping results in a substantial gradient of carrier concentration in the depletion layer with a built-in electric field.

Under reverse bias, the applied voltage increases both the electric field and depletion layer width (W), as shown in Fig. 3(b), but the carrier diffusion length (L) out of the depletion layer remains unchanged. Thus, more IR-generated carriers created in and near the depletion layer can contribute to the IR reverse current. On the other hand, heavy doping itself creates a lot of defects, contributing to the creation of many IR-generated carriers due to the sub-band-gap excitations.

As is known, impact-ionization occurs when a carrier gains sufficient energy from travelling in an electric-field to create an extra electron-hole pair. As this process continues it results in avalanche multiplication. Competing with impact-ionization process is energy loss by phonon emission [12-15].

Heavy doping results in more IR-generated carriers, raising impact-ionization rate under the influence of the built-in field. Thus, the impurity-photovoltaic-effect in the heavily doped diode enables avalanche effect to occur at lower voltage, creating breakdown. In this way, Zener breakdown is explained in terms of impurity-photovoltaic-effect.

Similar to a research work which discusses the temperature dependence of tunneling diode [6], this study discusses the temperature dependence of Zener breakdown as follows.

For a low reverse-bias voltage regime, Zener breakdown voltage decreases as the temperature is raised, since the number of IR-generated carriers in the diode increases with increasing temperature.

On the other hand, for a greater reverse-bias voltage regime, the breakdown voltage increases as the temperature is raised, since raising temperature increases phonon scattering. In this process, the breakdown voltage is related to an increase in carrier scattering at elevated temperatures, and phonon scattering plays an important role. As the temperature increases phonon scattering increases, shortening carrier mean free path and thereby reducing the ionization coefficients. Thus, raising temperature increases the breakdown voltage for the greater bias voltage regime [16,17].

If the doping is so heavy that a great many IR-generated carriers are created in the diode, the IR reverse current significantly increases with applied voltage even at low voltage range. In general, total reverse current dramatically increases with applied voltage.

In forward direction, the applied voltage decreases both W and built-in field, as shown in Fig. 3(c). Under the influence of substantial gradient of carrier concentration in the very heavily doped p-n junction diode, a strong forward diffuse current occurs. The forward diffuse current of the carriers significantly surpasses the reverse drift current of the IR-generated carriers. Therefore, total forward current dramatically increases with voltage.

According to the above mechanism based on impurity-photovoltaic-effect, the rectifying behavior of a p-n junction is lost under the condition of very heavy doping. Furthermore, the applied voltages make significant current flow occur in both forward and reverse current directions rather than in only one direction. Even at low applied voltages of either polarity, significant current flow occurs. Such p-n junction has a small contact resistance in both current directions, creating an ohmic contact.

In principle, the impurity-photovoltaic-effect applies to the explanation for heavily doped p-n junctions in all kinds of semiconductor materials.

## Ⅳ. CONCLUSIONS

In general, both Zener diode and ohmic contact resulting from a heavily doped p-n junction can be indeed explained by using the mechanism based on impurity-photovoltaic-effect in particle description. Not only quantum mechanics, but also the photovoltaic mechanism in particle description can explain these devices. This study provides a reasonable support for the mechanism of impurity-photovoltaic-effect due to IR absorption. Quantum mechanics is a great discipline and provides an effective description of electrons in view of wave. Considering wave-particle duality, the combination of quantum-mechanical tunneling and impurity-photovoltaic-effect mechanisms is suggested to explain the above devices. Each of the two mechanisms plays a big or small role in the explanation.

## DECLARATION OF COMPETING INTEREST

This is to certify that this manuscript does not make any conflict of interest with any person or institution or laboratories or any work.

## ACKNOWLEDGEMENT

This work was supported by Chinese Academy of Sciences (CAS).

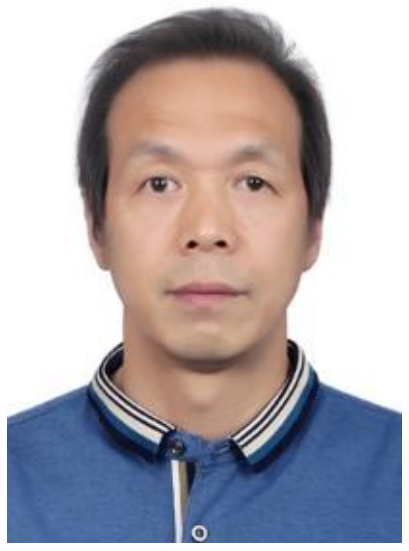

**Jianming Li**, research associate and senior engineer at the Institute of Semiconductors, Chinese Academy of Sciences. Email: jml_iscas@163.com; Tel.: +86-18310559676
Jianming Li was born in Beijing, China, in 1961. He studied at Beijing Normal University from 1980 to 1987. After graduated from the University, he joined the Institute of Semiconductors, Chinese Academy of Sciences. Since 1987, he has been engaged in research on photovoltaics, semiconductor materials and device physics, etc. As a Visiting Scientist, he worked at Brookhaven National Laboratory, USA, from 1992 to 1994. As a Visiting Scientist, he worked at Pennsylvania State University, USA in 1997.